\documentstyle[prd,aps,floats,graphicx]{revtex}

\begin{document}
\preprint{ astro-ph/0105253}
\draft
%
%
\input epsf
\renewcommand{\topfraction}{0.99}
\twocolumn[\hsize\textwidth\columnwidth\hsize\csname 
@twocolumnfalse\endcsname

\title{Viability of primordial black holes as short period gamma-ray
bursts
}
\author{Anne M.~Green}
\address{Astronomy Unit, School of Mathematical Sciences, Queen Mary 
University of London,\\ Mile End Road, London, E1 4NS,~~U.~~K.}
\date{\today} 
\maketitle
\begin{abstract}
It has been proposed that the short period gamma-ray bursts, which
occur at a rate of $\sim 10 {\rm yr^{-1}}$, may be evaporating
primordial black holes (PBHs). Calculations of the present PBH
evaporation rate have traditionally assumed that the PBH mass function
varies as $M_{{\rm BH}}^{-5/2}$. This mass function only arises if the
density perturbations from which the PBHs form have a scale invariant
power spectrum. It is now known that for a scale invariant power
spectrum, normalised to COBE on large scales, the PBH density is
completely negligible, so that this mass function is cosmologically
irrelevant. For non-scale-invariant power spectra, if all PBHs which
form at given epoch have a fixed mass then the PBH mass function is
sharply peaked around that mass, whilst if the PBH mass depends on the
size of the density perturbation from which it forms, as is expected
when critical phenomena are taken into account, then the PBH mass
function will be far broader than $ M_{{\rm BH}}^{-5/2}$.  In this
paper we calculate the present day PBH evaporation rate, using
constraints from the diffuse gamma-ray background, for both of these
mass functions. If the PBH mass function has significant finite width,
as recent numerical simulations suggest, then it is not possible to
produce a present day PBH evaporation rate comparable with the
observed short period gamma-ray burst rate. This could also have
implications for other attempts to detect evaporating PBHs.
\end{abstract}

\pacs{PACS numbers: 98.80Cq, 98.70 Rz   \hspace*{6.0cm}  astro-ph/0105253}

\vskip2pc]

\section{Introduction}
The physical origin of gamma-ray bursts (GRBs) is one of the
outstanding problems in astrophysics~\cite{grbrev}.  The majority of
gamma-ray burst are now known to have cosmological origin, however
Cline and co-authors have studied the short (duration $<200$ms) GRB
population~\cite{short}, and suggest that they may be due to the
evaporation of primordial black holes (PBHs) located in the galactic
halo~\cite{cline}. These short events, which make up roughly $2 \%$ of
the total population, have simple time histories and hard spectra,
relative to the longer duration GRBs, and are consistent with a
Euclidean source distribution, suggesting a local origin~\cite{cline}.

Primordial black holes (PBHs) are black holes which form in the early
universe~\cite{PBHform}. There are a number of possible mechanisms for
their formation, including the collapse of cosmic string
loops~\cite{cs} and the collision of bubbles formed at phase
transitions~\cite{pt}. The most natural formation mechanism is the
collapse of large inflationary density
perturbations~\cite{cl,pbhinf1,GL1,pbhinf2}. PBHs evaporate via the
emission of Hawking radiation~\cite{Hawking} and PBHs with mass $
M_{{\rm BH}} = M_{\star} \sim 5 \times 10^{14}$ g would be evaporating
today. In the standard picture of PBH evaporation~\cite{evappage} all
particles with rest mass less than the black hole temperature $T_{{\rm
BH}}$ where
\begin{equation}
T_{{\rm BH}} = \frac{ \hbar c^3}{8 \pi G M_{{\rm BH}}} = 1.06 \left( 
          \frac{10^{13} {\rm g}}
      {M_{{\rm BH}}} \right) {\rm GeV}\,,
\end{equation}
and $M_{{\rm BH}}$ is the PBH mass in grams, are emitted. The rate of
mass loss therefore depends on the number of particle degrees of
freedom, $\alpha(M)$:
\begin{equation}
\label{evaprate}
\frac{{\rm d} M_{{\rm BH}}}{{\rm d} t} = - \frac{ \alpha(M_{{\rm
              BH}})}{M_{{\rm BH}} ^2} \,.
\end{equation}
For the standard model of particle physics $\alpha(M_{{\rm BH}}) \geq
7.8 \times 10^{26} \, {\rm g^{3} \, s^{-1}}$~\cite{Hawking,conc2} for
$M_{{\rm BH}} \sim 5 \times 10^{14}$g. PBHs of this mass, which are
evaporating today, would have $T_{{\rm BH}} > \, 20 $MeV in the final
stages of their evaporation. Cline and collaborators argue that
$\alpha(M_{{\rm BH}})$ could then be significantly increased by the
effective number of degrees of freedom due to the quark-gluon phase
transition~\cite{cline}, resulting in a burst of radiation with
duration consistent with the observed duration of the short period
GRBs.

To calculate the PBH evaporation rate the PBH mass function and local
density enhancement are needed.  Cline and collaborators take the
bound on the global number density of PBHs per logarithmic mass
interval,  ${\cal N}_{{\rm g}} $, at $M=M_{\star}$, from the diffuse
gamma-ray background~\cite{evappage} to be $\leq 10^{5} \, {\rm
pc}^{-3}$. They combine this with a local density enhancement factor
$\eta= \rho_{{\rm l}}/ \rho_{{\rm g}} = 5 \times 10^{5}$, where `l'
and `g' denote local and global values respectively, to obtain the
local number density of PBHs per logarithmic mass interval at
$M=M_{\star}$, ${\cal N}_{{\rm l}} = \eta {\cal N}_{{\rm g}} \, \sim
10^{10} {\rm pc}^{-3}$.  The local PBH evaporation rate is then given
by
\begin{equation}
\frac{{\rm d} n_{{\rm BH}}}{{\rm d} t} = \frac{{\rm d} n_{{\rm BH}}}
         {{\rm d} M_{{\rm BH}}} \frac{{\rm d} M_{{\rm BH}}}{{\rm d} t} \,,
\end{equation}
which, using eq.~(\ref{evaprate}) and the fact that evaporating PBHs
can only be detected within $\sim 1 {\rm pc}$~\cite{cline}, becomes
\begin{equation}
\frac{{\rm d} n_{{\rm BH}}}{{\rm d} t}= \frac{
\alpha(M_{\star})}{M^{3}_{\star}} {\cal N}_{{\rm l}} \approx 10 \,
{{\rm yr}}^{-1} \,.
\end{equation}
The constraint on the global PBH number density ${\cal N}_{{\rm g}}=
{\rm d} n_{{\rm BH}} / {\rm d} \ln M_{{\rm BH}}$ has traditionally
been calculated~\cite{conc1,conc2} by assuming that the PBH mass
function is a power law $ {\rm d} n_{{\rm BH}} / {\rm d} M_{{\rm BH}}
\propto M_{{\rm BH}}^{-5/2}$ in which case the initial PBH number
density is given by~\cite{carr}
\begin{equation}
{\cal N}_{{\rm g}} = \frac{ \Omega_{{\rm PBH, 0}} \, 
                   \rho_{{\rm c}}}{2 M_{\star}} \,,
\end{equation}
where $\Omega_{{\rm PBH, 0}}$ is the present day fraction of the
critical energy density, $\rho_{{\rm c}}$ in PBHs. Using MacGibbon
and Carr's 1991 evaluation of the diffuse gamma-ray bound (which assumes
the $M^{-5/2}$ mass function)~\cite{conc1}:
\begin{equation}
\Omega_{{\rm PBH}}  \leq 7.6 (\pm 2.6) \times 10^{-9} h^{-1.95 \pm 0.15} \,,
\end{equation}
where $h$ is the Hubble parameters in units of $100 \, {\rm km s^{-1}
\, Mpc^{-1}}$, gives ${\cal N}_{{\rm g}} = 4.2 \times 10^{3} {\rm
\, pc^{-3}}$.

The power law PBH mass function, ${\rm d} n_{{\rm BH}}/ {\rm d}
M_{{\rm BH}} \propto M_{{\rm BH}}^{-5/2}$, was derived by Carr in the
1970s~\cite{carr} for scale-invariant density perturbations. In this
paper we outline the arguments which show that this mass function is
cosmologically irrelevant. We then calculate the constraint on the
global PBH evaporation rate for the sharply peaked mass function which
arises from assuming that all PBHs which form at a given epoch have
the same mass, and also for the far broader mass function found by
recent numerical studies~\cite{nj}, due to near critical gravitational
collapse~\cite{crit}.  Finally we review the calculation of the local
density enhancement factor.

\section{PBH mass function}
Early studies by Carr~\cite{carr}, which assumed that all PBHs which
form at a given epoch have the same mass ($M_{{\rm BH}} \approx
\gamma_{{\rm s}}^{3/2} M_{{\rm H}}$ where $M_{{\rm H}}$ is the horizon
mass at that epoch and $\gamma_{{\rm s}}$ parameterises the equation
of state: $p=\gamma_{{\rm s}} \rho$), found that an extended PBH mass
function is only possible if the primordial power spectrum is
scale-invariant (equal power on all scales). For Gaussian
distributed fluctuations the probability distribution of the smoothed
density field $p(\delta(M_{{\rm H}}))$ is given by
\begin{eqnarray}
p(\delta(M_{{\rm H}})) \, {\rm d} \delta(M_{{\rm H}})& =& \frac{1}{ 
         \sqrt{2 \pi} \sigma(M_{{\rm H}})}  \nonumber \\
     &&	\times \exp{\left( - \frac{\delta^2(M_{{\rm H}})}
        {2 \sigma^2(M_{{\rm H}})}\right)} \, 
	{\rm d} \delta(M_{{\rm H}}) \,,
\end{eqnarray}
where $\sigma(M_{{\rm H}})$ is the mass variance evaluated at horizon
crossing~\cite{ll}. For power law power spectra, $P(k) \propto k^{n}$,
where $P(k) = \langle |\delta_{{\bf k}} |^2 \rangle$ and $n$ is the
spectral index,
\begin{equation}
\label{sigma}
\sigma(M_{{\rm H}}) = \sigma(M_{{\rm H, 0}}) \left( \frac{M_{{\rm 
          eq}}}{M_{0}} \right)^{(1-n)/6} \left( \frac{ M_{{\rm H}}}
         {M_{{\rm eq}}} \right)^{(1-n)/4} \,,
\end{equation}
where `0' and `eq' denote quantities evaluated at the present day and
matter--radiation equality, and $\sigma(M_{{\rm H, 0}}) =9.5 \times
10^{-5}$~\cite{GL1} using the COBE normalisation~\cite{COBE}.  Using
the Press-Schechter formalism Kim and Lee found the initial mass
function produced by a power law power 
spectrum~\cite{kl}:
\begin{eqnarray}
\label{mf1}
\frac{{\rm d} n_{{\rm BH}}}{{\rm d} M_{{\rm BH}}} &=& \frac{n +3}{4}
       \sqrt{\frac{2}{\pi}} \gamma_{{\rm s}}^{7/4} \rho_{{\rm i}}
       M_{{\rm Hi}}^{1/2} M_{{\rm BH}}^{-5/2} \sigma_{{\rm H}}^{-1}
       \nonumber \\ &&  \times \exp{\left(-\frac{\gamma^{2}} {2
       \sigma_{{H}}^2} \right)} \,,
\end{eqnarray}
where $\rho_{i}$ and $M_{{\rm H i}}$ are the energy density and horizon
mass when the PBHs form, immediately after reheating finishes at the end
of inflation and the universe becomes radiation dominated. The horizon
provides a sharp lower cut-off in the mass function at $M_{{\rm BH}} = 
\gamma_{{\rm s}}^{3/2} M_{{\rm H i}}$.

For $n \neq 1$ the exponential term cuts the mass function off sharply
so that it is peaked around $M_{{\rm BH}} \sim M_{{\rm H i}}$. Only if
$n=1$ does the mass function have the traditionally used $M_{{\rm
BH}}^{-5/2}$ form.  However for a scale invariant power spectrum
normalised to COBE on large scales the PBH density is completely
negligible~\cite{cl,dgr}; the fraction of the energy density of the
universe in PBHs at the time they form is given by~\cite{carr}
\begin{equation}
\frac{\rho_{{\rm PBH}}}{\rho_{{\rm tot}}} \approx \sigma(M_{{\rm H}})
           \exp{ \left( -\frac{1}{18 \, \sigma^2(M_{{\rm H}})} \right)} \,.
\end{equation}
If $n=1$ then $\sigma(M_{{\rm H}}) =\sigma(M_{{\rm H, 0}})= 9.5 \times
10^{-5}$ and $\beta \sim 10^{(-6 \times 10^{6})}$! The $M^{-5/2}$ PBH
mass function is therefore cosmologically irrelevant. To produce an
interesting density of PBHs either a power-law power-spectrum with
more power on short scales ($n>1$), or a spike in the density
perturbations spectrum is required~\cite{cl,GL2}. In both of these
cases {\em if} all PBHs which form at a given epoch have the same mass
then the mass function is very sharply peaked.

Niemeyer and Jedamzik~\cite{nj} have shown using numerical simulations
that, as a consequence of near critical gravitational
collapse~\cite{crit}, at a fixed epoch PBHs with a range of masses
form. The PBH mass is determined by the size of the fluctuation from
which it formed:
\begin{equation}
\label{mbh}
M_{{\rm BH}}= k M_{{\rm H}} (\delta-\delta_{{\rm c}})^{\gamma} \,,
\end{equation}
where $\gamma$, $k$ and $\delta_{{\rm c}}$ are
constant for a given perturbation shape (for Mexican Hat shaped
fluctuations $\gamma=0.36$, $k=2.85$ and $\delta_{{\rm c}}=0.67$.)

\begin{figure}[t]
\centering
\leavevmode\epsfysize=6.3cm \epsfbox{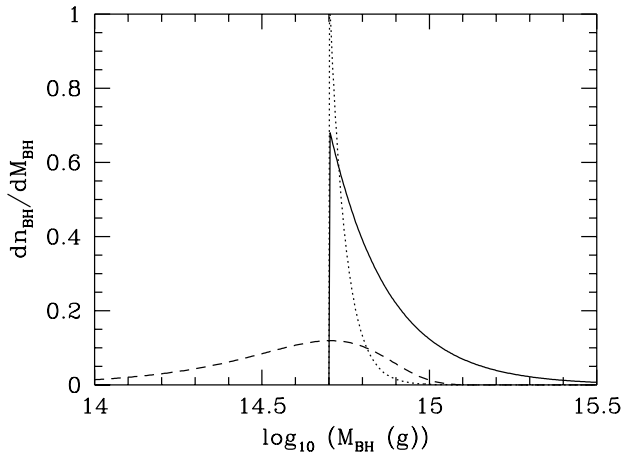}\\
\caption[mbhfig1]{\label{mbhfig1} The PBH initial number density, as a
function of mass, for the Carr $ M_{{\rm BH}}^{-5/2}$ (solid line),
Kim-Lee (dotted) and Niemeyer-Jedamzik (dashed) mass functions with
parameters chosen such that the present day densities, ignoring
evaporation, are the same ($\Omega_{{\rm PBH}}=1 \times 10^{-8}$). For
clarity the Carr and Niemeyer-Jedamzik mass functions are multiplied
by a factor of ten.}
\end{figure}

\begin{figure}[t]
\centering
\leavevmode\epsfysize=6.3cm \epsfbox{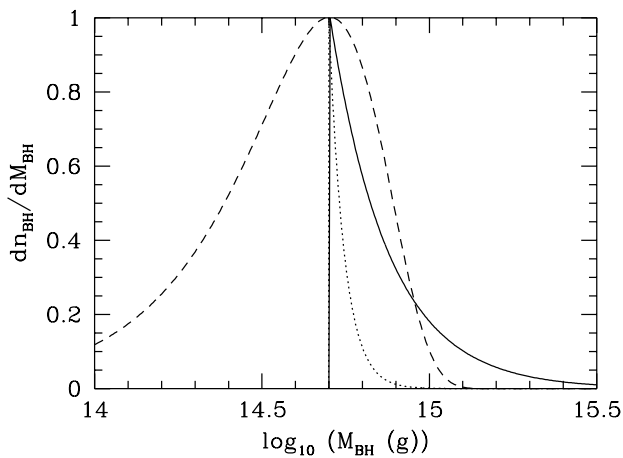} \\
\caption[mbhfig2]{\label{mbhfig2} The present day PBH number density,
as a function of mass, for the Carr $M_{{\rm BH}}^{-5/2}$ (solid
line), Kim-Lee (dotted) and Niemeyer-Jedamzik (dashed) mass functions
with parameters chosen such that the present day number density of
evaporating PBHs is ${\cal N}_{{\rm g}} = 4 \times 10^{3} {\rm
pc}^{-3}$.}
\end{figure}

It has been found that, for both power-law power spectra and flat
spectra with a spike on a particular scale, in the limit where the number
of PBHs formed is small enough to satisfy the observational
constraints on their abundance, it can be assumed that all the PBHs
form at a single horizon mass~\cite{GL2}. It is therefore possible to
calculate the PBH initial number density per unit mass interval
analytically~\cite{GL2,klr}:
\begin{eqnarray}
\label{mfnj}
\frac{{\rm d} n_{{\rm BH}}}{{\rm d} M_{{\rm BH}}} & = & 
       \frac{\rho_{i}}{\sqrt{2 \pi}  \gamma \sigma(M_{{\rm H}}) M_{{\rm BH}} M_{{\rm H}}}
       \left( \frac{M_{{\rm BH}}}{kM_{{\rm H}}}\right)^{1/\gamma}
        \nonumber \\ &&
      \exp{ \left\{ -\frac{ \left[\delta_{{\rm c}} + (M_{{\rm BH}}/ 
      k M_{{\rm H}})^{1/\gamma} \right]^2}{2 \sigma^2(M_{{\rm H}})} \right\} }
           \,,
\end{eqnarray}
where $\sigma(M_{{\rm H}})$ is defined in Eq.~(\ref{sigma}). We
subsequently refer to this mass function as the Niemeyer-Jedamzik mass
function. The physical number density of PBHs dilutes $\propto
a^{-3}$ so that at any later time t
\begin{equation}
\frac{{\rm d} n_{{\rm BH}}}{{\rm d} M_{{\rm BH}}}(t)=\frac{{\rm d} n_{{\rm BH}}}{{\rm d} M_{{\rm BH}}} \left( \frac{T(t)}{T_{{\rm i}}} \right)^{3} \,,
\end{equation}
where $T_{{\rm i}}$ can be related to $M_{{\rm H i}}$ using~\cite{GL1}
\begin{equation}
M_{{\rm H i}}= M_{{\rm H 0}} \left( \frac{ T_{{\rm eq}}}{T_{{\rm i}}}
 \right)^{2} \left( \frac{T_{0}}{T_{{\rm eq}}} \right)^{3/2} \,.
\end{equation}

The three initial mass functions (Carr $ M_{{\rm BH}}^{-5/2}$, Kim-Lee
and Niemeyer-Jedamzik) are plotted in Fig.~\ref{mbhfig1} with the
parameters of each mass function chosen such that the energy density
in PBHs is the same, corresponding to a present day PBH energy density
of $\Omega_{{\rm PBH}} = 1 \times 10^{-8}$~\footnote{Strictly speaking
in the case of the Niemeyer-Jedamzik mass function the present day
density will be less than this as the low mass tail of PBHs with
$M_{{\rm BH}}< 5 \times 10^{14}{\rm g}$ will have evaporated}. In
Fig.~\ref{mbhfig2} we plot the present day mass functions with
parameters chosen such that ${\cal N}_{{\rm g}}= 4 \times 10^{3} {\rm
pc}^{-3}$. The Kim-Lee mass function, which arises from assuming that
all PBHs which form at a given epoch have the same mass, is very
sharply peaked. The Niemeyer-Jedamzik is far broader, with a long tail
of low mass PBHs which would have evaporated since $ z \sim 700$ and
would contribute to the diffuse gamma-ray background.

\section{PBH evaporation rate}

The abundance of PBHs evaporating today is constrained by their
contribution to the diffuse gamma-ray
background~\cite{evappage,conc1,y,dgr,klr}.  All PBHs which have
evaporated since $z=700$ contribute to the diffuse gamma-ray
background~\cite{700}, therefore the resulting constraint on the
number of PBHs evaporating today depends on the PBH mass function; the
broader the mass function the tighter the constraint on the number of
PBHs evaporating today. The diffuse gamma-ray bound has been
recalculated using recent measurements by COMPTEL~\cite{COMPTEL} and
EGRET~\cite{EGRET} for the Kim-Lee mass function~\cite{dgr} and also
for the Niemeyer-Jedamzik mass function~\cite{klr}.

Since the Kim-Lee and Carr $M_{{\rm BH}}^{-5/2}$ mass function both
have a sharp lower cut-off, the diffuse gamma-ray constraint on the
present day PBH evaporation rate is the same for both mass
functions~\footnote{The resulting constraint on the mass density in
PBHs of all masses, $\Omega_{{\rm PBH}}$, is far tighter for the
Kim-Lee mass function however, since that mass function is far more
sharply peaked.} , with the recent COMPTEL and EGRET measurements
tightening the constraints on ${\cal N}_{{\rm g}}$ by a factor of
about 1.5~\cite{CMnew}. For the Niemeyer-Jedamzik mass function the
relevant constraint on PBHs with mass $M=M_{\star}$ is $\sigma(M_{{\rm
H}}=1.54 \times 10^{15} {\rm g}) < 0.056$.  Inserting this in
Eq.~(\ref{mfnj}) leads to ${\cal N}_{{\rm g}} < 2.7 \times 10^{-4}
h^{2} {\rm pc^{-3}}$ so that
\begin{equation}
\frac{{\rm d} n_{{\rm BH}}}{{\rm d} t}= \frac{
\alpha(M_{\star})}{M^{3}_{\star}} {\cal N}_{{\rm l}} \approx 
5.2 \times 10^{-14} h^2 \eta \, {{\rm pc^{-3}}} \,
{{\rm yr}}^{-1} \,.
\end{equation}
This calculation assumes that the cosmological constant is zero,
$\Omega_{\Lambda}=0$. If $\Omega_{\Lambda}=0.7$, as current
observation appear to indicate~\cite{lam}, then the constraint is
tightened to $\sigma(M_{\star}) < 0.023$~\cite{klr} and ${\cal
N}_{{\rm g}} < 4.5 \times 10^{-157} h^{2} {\rm pc^{-3}} $ so that the
present day rate of PBH evaporation is completely negligible.

We caution that the constraints on ${\cal N}_{{\rm g}}$ are very
sensitive to small changes in the limits on $\sigma(M_{{\rm H}})$ (or
equivalently $n$) due to the exponential factor which arises in the
expressions for ${\rm d} n_{{\rm BH}}/{\rm d} M_{{\rm BH}}$
(Eqs.~(\ref{mf1}) and (\ref{mfnj})). This illustrates the fact that to
produce an interesting (i.e. non-negligible) density of PBHs
(evaporating today or otherwise) requires fine tuning of the size of
the density perturbations from which the they form. We have also seen
that if the PBH mass function has significant finite width then the
diffuse gamma-ray bound (which constrains the abundance of all PBHs
which have evaporated since a redshift of $z=700$) places far tighter
limits on the present day rate of PBH evaporation than if the PBH mass
function is sharply peaked.

\section{PBH concentration}
The local density enhancement factor is usually
calculated~\cite{conc1} by assuming an isothermal halo so that the
density at galactocentric radius $R$ (outside the core radius $R_{{\rm
c}}$) is related to the asymptotic circular velocity $V_{\infty}
\approx 220 {{\rm km s^{-1}}}$ by~\cite{cl87}
\begin{eqnarray}
\rho_{{\rm h}}(R)&=& \frac{V_{\infty}^2}{ 4 \pi G R^2} \nonumber \\ 
 & \approx & 6.1 \times 10^{25} \left( \frac{V_{\infty}}{220 {\rm km s^{-1}}}
\right)^2 \left( \frac{R}{10 {\rm kpc}} \right)^{-2} {{\rm g cm^{-3}}}
\,.  
\end{eqnarray} 
Taking our galactocentric radius to be
$R_{\odot}=8.5 (\pm 1.1){\rm kpc}$ gives a local density enhancement factor
\begin{equation}
\eta \approx 4.5 (\pm 0.6) \times 10^{5} h^{-2} \left( \frac{\Omega_{{\rm h}}}
                 {0.1} \right)^{-1} \,,
\end{equation}
where $\Omega_{{\rm h}}$ is the fraction of the critical density in
galactic halos. Halzen et. al.~\cite{conc2} claim, without explicit
calculation, a larger value for the local density enhancement:
\begin{equation}
\eta \approx 1.36 (\pm 0.9) \times 10^{7} h^{-2} \left( \frac{\Omega_{{\rm h}}}
                 {0.1} \right)^{-1} \,,
\end{equation}
which apparently~\cite{wright} relies on the assumption that PBHs are
concentrated to the same extent as luminous matter.  

An alternative procedure is to calculate $\eta$ directly from
estimates of the local halo density. The local halo density is poorly
known, with the most recent estimates giving $ \rho_{{\rm l}}=
9.2^{+3.8}_{-3.1} \times 10^{-25} \, {\rm g \,
cm^{-3}}$~\cite{GGT}. Taking the currently favoured value of the
present day mass density $\Omega_{{\rm m}}=0.3$~\cite{lam} gives
$\rho_{{\rm m}} \approx 5.6 \times 10^{-30} h^{-2} \, {\rm g
\,cm^{-3}} $. Combining these values produces
\begin{equation}
\eta=1.6^{+1.8}_{-0.8} \times 10^{5} \, h^{-2} \left( \frac{\Omega_{{\rm m}}}
                 {0.3} \right)^{-1}\,,
\end{equation}
which is in broad agreement with the value calculated from the
asymptotic circular velocity assuming that the halo is isothermal.
Since evaporating PBHs could only be detected by BATSE out to
distances of order a parsec~\cite{cline}, it appears that, if the PBH
mass function has significant width as recent numerical simulations
suggest, then the local density enhancement factor is not large enough
to produce a local PBH evaporation rate comparable with the observed
frequency of short period gamma-ray bursts ($10-20 \, {\rm yr^{-1}}$).

\section{Conclusions}
Calculations of the present PBH evaporation rate have traditionally
assumed that the PBH mass function varies $\sim M_{{\rm
BH}}^{-5/2}$. This mass function only arises if the density
perturbations from which the PBHs form have a scale invariant power
spectrum, in which case the PBH density is completely
negligible~\cite{cl,dgr}. We have recalculated the present PBH
evaporation rate, using the bounds which arise from the diffuse
gamma-ray background~\cite{dgr,klr}, for the sharply peaked mass
function which arises if all PBHs which form at a given epoch have
the same mass and also for the broader PBH mass function found by
recent numerical simulations. We find that if the PBH mass function
has significant finite width it is not possible to produce a present
day PBH evaporation rate comparable with the observed short period
gamma-ray burst rate. This could also have implications for other
attempts to detect evaporating PBHs.

\section*{Acknowledgements}

A.M.G.~was supported by PPARC and acknowledges use of the Starlink
computer system at Queen Mary, University of London. 


\begin{references}
\bibitem{grbrev} P. Meszaros, Nucl. Phys. Proc. Suppl. 80, 63 (2000).
\bibitem{short} C. Barat et. al. Astrophys. J {\bf 285}, 791 (1984).
\bibitem{cline} D. B. Cline and  W. P. Hong, Astrophys. J 
             {\bf 401}, L57 (1992);
              D. B. Cline, Nucl. Phys. A {\bf 610}, 500 (1996);  
              D. B. Cline, D. A. Sanders and W. P. Hong, Astrophys. 
              J {\bf 486}, 169
               (1997); D. B. Cline, C. Matthey and S. Otwinowski, 
                 Astrophys. J 
              {\bf 527}, 827 (1999); D. B. Cline, C. Matthey and S. 
               Otwinowski, astro-ph/0105059.
\bibitem{PBHform} Ya. B. Zeldovich and I. D. Novikov, Sov. Astron. A. J. 
                  {\bf 10}, 602 (1967); S. W. Hawking, Mon. Not. Roy. Astron.
                  Soc. {\bf 152}, 75 (1971); B. J. Carr and S. W. Hawking,
                   Mon. Not. Roy. Astron. Soc {\bf 168}, 399 (1974).
\bibitem{cs} A. G. Polnarev and R. Zembowicz, Astron. Zh. {\bf 58}, 706
              (1988); S. W. Hawking, Phys. Lett. B {\bf 231}, 237 (1989).
\bibitem{pt} S. W. Hawking, I. Moss and J. Stewart, Phys. Rev. D {\bf 26},
            2681 (1981); D. La and P. J. Steinhardt, Phys. Lett. B {\bf 220},
            375 (1989).
\bibitem{cl} B. J. Carr and J. E. Lidsey, Phys. Rev. D {\bf 48}, 543
(1993).
\bibitem{pbhinf1} B. J. Carr, J. H. Gilbert and J. E. Lidsey, Phys.
               Rev. D {\bf 50}, 4853 (1994);
               P. Ivanov, P. Naselsky and I. Novikov, Phys. Rev. D
               {\bf 50}, 7173 (1994); J. Garcia-Bellido, A. D. Linde
               and D. Wands, Phys. Rev. D {\bf 54}, 6040 (1996);
               J. Yokoyama, Astron. and Astrophys. {\bf 318}, 673 (1997).
\bibitem{GL1} A. M. Green and A. R. Liddle, Phys. Rev. D {\bf 54}, 
              6166 (1997).
\bibitem{pbhinf2} J. Yokoyama, Phys. Rev. D {\bf 58}, 083510 (1998);
               M. Kawasaki and T. Yanagida, Phys. Rev. D {\bf 59},
               043515 (1999).
 
\bibitem{Hawking} S. W. Hawking, Nature {\bf 30}, 248 (1974).
\bibitem{evappage} D. N. Page and S. W. Hawking, Astrophys. J, 
                {\bf 206}, (1976).
\bibitem{conc2} F. Halzen et. al. Nature {\bf 353}, 807 (1991).

\bibitem{conc1} J. H. MacGibbon and B. J. Carr, ApJ {\bf 371}, 447 (1991).
\bibitem{carr} B. J. Carr, Astrophys. J. {\bf 201}, 1 (1975).

\bibitem{nj} J. C. Niemeyer and K. Jedamzik, Phys. Rev. Lett. {\bf 80}, 5481
              (1998); J. C. Niemeyer and K. Jedamzik, Phys. Rev. D  {\bf 59}, 
              124013 (1999).
\bibitem{crit} M. W. Choptuik, Phys. Rev. Lett. {\bf 70}, 9 (1993).
\bibitem{ll} A. R. Liddle and D. H. Lyth, Phys. Rep. {\bf 231}, 1
         (1993).
\bibitem{COBE} E. F. Bunn, A. R. Liddle and M. White,
	Phys. Rev. D {\bf 54}, 5917 (1996); E. F. Bunn and
	M. White, Astrophys. J. {\bf 480}, 6 (1997).
\bibitem{kl} H. I. Kim and C. H. Lee, Phys. Rev. D {\bf 54}, 6001
(1996).  

\bibitem{dgr} H. I. Kim, C. H. Lee and J. H. MacGibbon, Phys. Rev. D {\bf 59},
             063004 (1999).
\bibitem{GL2} A. M. Green, A. R. Liddle, Phys. Rev. D {\bf 60}, 063509 (1999).
\bibitem{klr} G. D. Kribs, A. K. Leibovitch and I. Z. Rothstein, Phys. Rev.
           D {\bf 60}, 103510 (1999).
\bibitem{y} J. Yokoyama, Phys. Rev. D {\bf 58}, 107502 (1998). 
\bibitem{700} A. A. Zdiarski and R. Svensson, Astrophys. J. {\bf 344},
          551 (1989).

\bibitem{COMPTEL} S. C. Kappadath et. al., Astron. and Astrophys. Suppl.
        {\bf 120}, C619 (1996).
\bibitem{EGRET} P. Sreekumar et. al., Astrophys. J. {\bf 494} 523 (1998).
\bibitem{CMnew} B. J. Carr and J. H. MacGibbon, Phys. Rep. {\bf 307},
               141 (1998).
           
\bibitem{lam} M. Tegmark and M. Zaldarriaga, Astrophys. J. {\bf 544},
            30 (2000).
\bibitem{cl87} B. J. Carr and C. G. Lacey, Astrophys. J. {\bf 316}, 23 (1987).
\bibitem{wright} E. L. Wright, Astrophys. J. {\bf 459}, 487 (1996).             
\bibitem{GGT} E. I. Gates, G. Gyuk and M. S. Turner, Astrophys. J. {\bf 449},
        L133 (1995).









\end{references}
\end{document}